# *The need for structural changes to create impactful public engagement in US particle physics*

**Contact Information:**
Sarah Demers, Yale University, sarah.demers@yale.edu
Kathryn Jepsen, SLAC National Accelerator Laboratory, kjepsen@slac.stanford.edu
Don Lincoln, Fermi National Accelerator Laboratory, lincoln@fnal.gov
Azwinndini Muronga, Nelson Mandela University, Azwinndini.Muronga@mandela.ac.za

**Authors:**
Kétévi Assamagan, Brookhaven National Laboratory
Mateus Carneiro, Brookhaven National Laboratory
Sarah Demers, Yale University
Kathryn Jepsen, SLAC National Accelerator Laboratory
Don Lincoln, Fermi National Accelerator Laboratory
Azwinndini Muronga, Nelson Mandela University

## Executive summary:

This Snowmass21 Contributed Paper addresses the structural changes that need to occur in the many groups and organizations that intersect with the US particle physics community to enable impactful public engagement to flourish. The impetus for these changes should come from the particle physics community, which should acknowledge the importance of public engagement and act on the recommendations in this Snowmass contributed paper.

Scientists have expressed frustration at the barriers, penalties and lack of support that discourage them from participating in public engagement. In this paper, we provide many ways to create a supportive, enabling atmosphere for public engagement among physicists. These recommendations include:

- Providing or financially supporting training in effective public engagement
- Supporting the creation of public engagement programs that scientists can participate in
- Codifying the importance of public engagement in official documents such as:
  - Laboratory contracts
  - Faculty handbooks
  - Professional society strategic plans

- - Experimental collaboration constitutions
    - Merit criteria used by institutions that fund research
  - Considering public engagement along with activities such as service and teaching in:
    - Hiring
    - Tenure
    - Promotion
    - Other reviews
  - Funding public engagement work as part of grant proposals
  - Incorporating public engagement into conferences and meetings in the form of:
    - Plenary talks
    - Parallel sessions
    - Public lectures
    - Training opportunities for conference participants
    - Public engagement opportunities for conference participants
  - Recognizing and rewarding scientists who contribute to public engagement efforts
  - Encouraging others, including peers, mentees and students, by participating in and discussing the importance of public engagement

Many individuals, groups and organizations are already implementing recommendations like these. We encourage you to find the areas in which you can make structural and cultural changes to better enable scientists to participate in public engagement. It should be the goal of the US particle physics community and the stakeholders named in this document to implement these recommendations between now and the next Snowmass process.

# 1. Introduction
## 1.1 Motivation

Engaging the public is an important part of particle physics research. It informs and inspires the next generations of scientists, including those who have not traditionally had access to careers in physics or been encouraged to pursue careers in physics. It fulfills scientists' duty to share the benefits of their work with members of the public. It spurs scientists to think in a new way about their research and why they do it. It communicates the value of particle physics research to those who decide whether it is worth funding. Public engagement about particle physics is in itself a benefit of particle physics research.

But participating in public engagement can be difficult for scientists for a variety of structural and cultural reasons, as the authors of this paper found in the responses to a survey we conducted of members of the US particle physics community.

In this paper, we will describe the current situation, via those survey responses, and recommend changes in a variety of areas to better enable scientists to participate in public engagement. We will make recommendations for changes in the following areas:

- within the particle physics community—in research groups, experimental collaborations and at conferences;
- at universities and colleges—in departments, schools and colleges, and among leadership;
- at the national laboratories;
- in the Office of Science and Technology Policy and in Congress;
- at institutions that fund research, including the National Science Foundation, the Department of Energy and private foundations; and
- in professional bodies and societies, including the American Association for the Advancement of Science, the American Physical Society, and the APS Division of Particles and Fields.

In each section, we will discuss how public engagement by scientists is beneficial to the group or organization in focus. We will discuss how those groups and organizations can affect scientists' ability to participate in public engagement. We will then recommend steps those groups and organizations should take to better enable and support public engagement.

Not every scientist needs to contribute to public engagement in the same way. There are a variety of ways to participate, including pushing for the types of structural changes we recommend in this paper. We encourage you to read with an eye toward the areas in which you can have an influence.

# 1.2 Survey

In 2020, the Public Education & Outreach group conducted a survey of physicists, physics students, technicians, engineers and other staff who are connected to the US particle physics community. Of the respondents, 24 students, 47 postdocs, 45 non-tenured physicists, 99 tenured physicists and 5 other staff said they had participated in public engagement in the previous two years. Eight students, 19 postdocs, 7 non-tenured physicists, 25 tenured physicists and 6 other staff said they had not participated in public engagement in the previous two years.

The No. 1 reason given for not participating in public engagement was "time constraints/not enough time." The majority, about 78%, of the respondents, including those who did and did not participate in engagement, listed this as a barrier. The second most common response was "lack of/not enough benefit to career and funding opportunities." About 38% of respondents listed this as a barrier.

The Public Education & Outreach group considers these two roadblocks to be related, an idea that survey comments support. If public engagement were seen as an essential part of physicists' jobs, physicists would be better able to make time for it. If public engagement were recognized and rewarded in physicists' jobs, physicists would be encouraged to make it more of a priority.

Not only do physicists generally receive neither recognition nor reward for participating in public engagement, but survey respondents said they are often penalized for it. As one said: "It appears as though outreach is perceived as a 'distraction' from research, such that participating in outreach may jeopardize or have negative consequences on a physicist's promotion evaluation."

Another commented: "Academic physics is extremely competitive. If (especially junior) scientists aren't focusing on research, they won't survive." Another said they did outreach "out of 'the good of my heart' because I like our field and what we study, and I want to share that with others," but that "leaning on the 'goodness of scientists' to ensure that outreach/engagement is done is only sustainable at the sacrifice of burning out people by spreading them too thin."

Others pointed out that it is not feasible for all scientists to do public engagement on their own time, without compensation—and potentially at the cost of their careers. One respondent said, "you can argue we do it because we care, and for the good of the community," but "if I don't get tenure, community engagement won't pay my bills!"

Still another respondent said: "We need a system that rewards the time spent in science communication… Young folks cannot spend the time needed for training and efficient outreach if the field doesn't value outreach in terms of career opportunities. We should not be put in the position to choose between spending time engaging with the public/wider community and 'doing real science.'"

Many respondents called for training, provided to and easily accessible to all rather than only those who seek it out. One suggested: "Make it a more normal part of the physics experience so that people have the tools and comfort to be able to do more outreach." They explained that "[t]he view of my peers has generally been that 'outreach people' are not great researchers since they devote significant time and learning to do outreach that could otherwise be used for writing research papers, etc."

Multiple respondents pointed out the need for cultural change in the physics community. Echoing several other commenters, one respondent said that public engagement "needs to be a more explicit part of our roles." Another said: "The community needs to value it more, and start to see it as a part of a scientist's job description, and not just something especially motivated people do in their free time." Another commenter pointed out that teaching, unlike outreach, "is viewed as an obligation of many academic positions." They said: "We likely need a systemic expansion of what is traditionally 'teaching' to include these outreach activities."

Respondents said that they needed more options for getting involved in public engagement. One respondent said that outreach on large experiments "seems to be done only by a small handful of people, all of whom devote such a large portion of their time to it that it detracts from their physics work. It would need to become something to which many people can contribute a small amount instead, and it's not currently set up for that." Another respondent suggested "normalizing outreach events that are more modern, such as participating in Instagram takeovers." They said that in their experience, colleagues had "scoffed" at this type of engagement. Negative attitudes about public engagement can be contagious, another respondent pointed out; they said that they had seen professors transmit these views to their students.

Many others pointed out the need for structural change. As one respondent said: "Change should come from management recognition of outreach." Another explained that "management should encourage and create the conditions by which those engagements are beneficial to one's performance and career growth and development." Management could give physicists time to do public engagement by hiring sufficient staff to cover other work, one commenter suggested. Another suggested the idea of an "engagement holiday" to give people time to do engagement.

Many others suggested structural change needed to come at the level of funding agencies. As one respondent said: "A major cultural shift of appreciating the importance of these roles could only come from the funding agencies. The program manager who funds me is not interested in my time or my postdoc's time spent on science communication." One respondent who said they worked at a national laboratory said "it would help if education or public outreach (since we're funded with tax dollars) were explicitly supported with monies from the DOE and were part of the mission statement so that individuals could spend more time on science communication." Yet another said: "Grant applications/funding should specify a time or funding allotment dedicated to science communication, education or outreach," and another suggested that grant reviewers give "[e]xplicit credit for time spent on these activities."

In this paper, we follow the advice of these voices from the physics community, recommending changes at a variety of levels that, together, could better enable physicists to participate in public engagement.

# 2. Focus areas

## 2.1 Particle physics community

### 2.1.1 Research groups

**How public engagement helps research groups achieve their goals:**
Public engagement can raise the profile of a research group within an institution and more broadly within the research field. Public engagement can help a research group recruit new students and postdocs.

**Why research groups are influential in enabling scientists to participate in public engagement:**
As students and postdocs progress through their training in particle physics, much of their experience is shaped by their research mentors in the context of their research groups. PIs set the expectations, provide funding and opportunities, and demonstrate what is needed for success in the field through their actions. The research group is therefore an important site for the culture of particle physics to change such that the work of public engagement becomes a component of the job of active researchers in our field.

**Recommendations:**
Research mentors should ensure that the environments of their research groups are conducive to mentees participating in public engagement. Steps they can take include:
- Provide opportunities for mentees to receive training in various areas of public engagement. Encourage mentees to think of high-quality engagement as something that can be learned to do, rather than an innate skill.
- Provide mentees with public engagement opportunities.
- Set expectations within the research group that some time should be devoted to public engagement.
- Provide funding for public engagement, in the form of supporting travel and/or materials.
- Model the importance of public engagement by participating in this work.
- Discuss best practices and experiences with public engagement within the research group.
- Encourage sharing of skills and knowledge not only from the PI, but also allow knowledge to flow up from the trainees, who may bring their own experiences, modes of engagement, and access to audiences via different platforms.

# 2.1.2 Experimental collaborations

**How public engagement helps experimental collaborations achieve their goals:**
Experimental collaborations benefit from the public being aware of their research successes. Visibility enhances the ability of PIs to get new grants and maintain current funding levels.

**Why experimental collaborations are influential in enabling scientists to participate in public engagement:**
Experimental collaborations are extremely important in shaping the culture of particle physics. Collaborations can provide resources to support public engagement efforts; they can provide public engagement training to collaboration members; and they can include in their organizational structures positions dedicated to public engagement.

**Recommendations:**
- Each experiment should identify an individual or group of individuals who are willing to spend some of their time on science outreach. The collaboration should enact a tax on member institutions to support outreach or should identify funds directly originating from the funding agencies that should be earmarked to support public engagement.

- Experimental collaborations should codify the collaboration's commitment to outreach in their governing documents. This will allow public engagement to remain a priority through changes in experimental leadership.
- Experiments should include presentations on public engagement in plenary sessions of collaboration meetings.
- When experimental collaborations make public announcements, the individuals responsible for public engagement should be heavily involved.

# 2.1.3 Conferences

**How public engagement helps conferences achieve their goals:**
Conference organizers put together programs to benefit attendees. Public engagement training and opportunities to participate in public engagement are two benefits that conference organizers can offer. Public engagement can raise the profile of a recurring conference, making it more attractive to potential future participants.

Some institutions host conferences to raise their profiles among both attendees and their local communities. Receiving that recognition requires putting together a valuable program and making sure people know about it. Public engagement activities such as public lectures allow the local community to benefit from the conference. Engaging the public may also draw media attention to the institution and to the topics under discussion at the conference.

**Why conferences are influential in enabling scientists to participate in public engagement:**
A scientific conference gathers together a great number of researchers. This provides both an opportunity for the scientific community to address the local community and an opportunity to give scientists training and experience with public engagement. Participating in public engagement training or activities with colleagues could help build an atmosphere of support for such activities among members of the scientific community.

**Recommendations:**
- For each conference, there should be one or more public lectures or other public engagement activities, to be held on the weekend or evening so as to improve the chances of good attendance. In the case of public lectures, care must be taken to select a speaker with experience speaking with a general audience. Attendance will be larger if the lecture covers a topic that is interesting to a general audience.
- Conference organizers should engage local university or laboratory press officers, who are familiar with the potential audiences in the area, to help plan and advertise public engagement activities. Applications to host scientific conferences should include information about how this will occur and about any opportunities to provide public engagement training to conference attendees.
- Conference organizers should ensure that at least one parallel session is devoted to public engagement.

- Conference organizers should ensure that at least one plenary talk is devoted to public engagement.

# 2.2 Universities and colleges

**How public engagement helps colleges and universities achieve their goals:**
Colleges and universities can benefit tremendously from effective public engagement by their researchers. The local and national or international reputation of the institute can be enhanced, and students can be attracted to apply to institutions when they hear about exciting research occurring at the school. There is also an opportunity for public engagement to lead to stronger ties between campuses and their surrounding communities. Partnerships between institutions and their local communities can work in support of efforts to diversify our field, by broadening the base of people who have access to results in particle physics and who therefore might be inspired to pursue studies in our field or related disciplines.

**Why colleges and universities are influential in enabling scientists to participate in public engagement:**
The processes and cultures of colleges and universities have important roles to play in setting good conditions for public engagement activities to thrive. The rubrics for hiring and promotion within colleges and universities help set the priorities of the faculty, and therefore the directions in which PIs lead their research groups. If public engagement were expected, supported and rewarded, the culture within research groups would respond, and in turn the culture of particle physics would change.

**Recommendations:**
General:
- Institutions should consider mechanisms for allowing some fraction of faculty effort certification, or summer salary, to be charged to public engagement accounts.

At the hiring stage:
- Applicants should be encouraged to include statements regarding their public engagement experiences and plans, and to highlight this work on their CVs.
- Colleges and universities should include Interview questions about public engagement. This would send a message that the department values these opportunities and would allow institutions to hire faculty with strength in this area.

At the review stage:
- Faculty should be encouraged or required to include a statement describing their outreach activities in their materials. Public engagement could be explicitly included in the "service" or "teaching" categories, depending on the nature of the work.
- Those reviewing should consider public engagement in review- and promotion-related decisions.

- Best practices and metrics of success in engagement should be explicitly expected in the review and promotion process.
- When local experts are not available in particular engagement areas, external experts could be invited to provide letters of evaluation or support as part of the formal review process.

# 2.2.1 Departments, schools and colleges

**How public engagement helps departments, schools, and colleges achieve their goals:**
Just like colleges and universities, specific departments and schools can benefit tremendously from effective public engagement by their researchers. Public engagement can develop their reputations and attract new students.

**Why departments, schools, and colleges are influential in enabling scientists to participate in public engagement:**
Faculty members depend on, and are at the mercy of, their department colleagues throughout their careers, due to tenure and promotion decisions, committee and teaching assignments, and the general collegial atmosphere (or lack thereof) in which they need to function. The opinions and values of department members and chairs can therefore strongly influence the actions that faculty members take. When a department, from the chair to other members of the department, recognizes the importance of public engagement, this work becomes much more feasible.

Unfortunately, in many cases, the work of public engagement can be seen as a distraction from a primary mission of research by faculty in physics departments. Participation in public engagement can even be viewed negatively, as an indicator of a lack of seriousness or rigor on the part of the scientist. A shift to a supportive structure for public engagement within physics departments could make a significant impact on the willingness of researchers to engage in this work.

**Recommendations**:
- Formally incorporate recognition of public engagement in hiring, tenure and promotion decisions, by considering public engagement efforts either on their own or in conjunction with service and teaching.
- Formally incorporate leadership in the area of public engagement in the process of choosing the department chair.
- Codify these requirements in the faculty handbook.
- Financially support efforts to provide scientists training in public engagement.

# 2.2.2 University leadership

**How public engagement helps university leadership achieve their goals:**
Good public engagement can raise the profiles of both the individuals doing the work and the institutions they represent.

**Why university leadership is influential in enabling scientists to participate in public engagement:**
The tone is often set at the top, and institutional leadership at colleges and universities have an important role to play in encouraging public engagement among their students and faculty. University leadership can provide a significant boost to public engagement efforts, as the reach of campus media typically extends far beyond what an individual researcher can access on their own.

A true cultural shift to take public engagement seriously depends on the recognition of the work in hiring and promotions on campus. The campus leadership plays an important role in ensuring that this work is considered in hiring and promotion on their campuses.

**Recommendations:**
- Formally incorporate recognition of public engagement in hiring, tenure and promotion decisions, by considering public engagement efforts either on their own or in conjunction with service and teaching.
- Highlight public engagement efforts externally, in communications with alumni and the public.
- Highlight public engagement efforts internally, in communication with campus communities.
- Financially support efforts to provide scientists training in public engagement.
- Financially support public engagement efforts.

## 2.3 National labs

**Background:**
National laboratories are a complex ecosystem, incorporating many interlocking interests. The primary groups are the laboratory as an institution and the scientific community, broadly defined. Within the scientific community, there are further subdivisions, for example the scientists and technical staff, or the laboratory employees and visitors who use the facility. In addition, in many laboratories there exist collaborations of individuals who have banded together to conduct an experiment.

Laboratory management has multiple layers. Senior laboratory management can influence middle management, which can influence individual scientists and technical staff.

Institutional communication by the laboratory to the public is handled by communication professionals. This communication can be buttressed by laboratory scientists via conversations with the media, OpEds, videos, etc. Collaborations usually work with their respective laboratory's management and professional communication staff to develop and release communications such as press releases describing scientific advances.

Outside of the strictures of formal laboratory communication and press releases, some individual scientists and technical professionals choose to participate in public engagement of their own volition. Such outreach might include speaking with local community groups, writing OpEds, creating content suitable for social media, interacting with reporters, visiting schools, etc. Many of these activities occur outside of normal laboratory operating hours.

**How public engagement helps national labs achieve their goals:**
Laboratories rely on the support of their local communities to be able to take up space and construct new facilities. In their contracts with the Department of Energy, laboratories agree to "[b]uild a positive, supportive relationship founded on openness and trust with the community and region in which the Laboratory is located." (For example, see Argonne's contract.) An article on "Community, Science, and Your Neighbor" published by DOE mentions that this can be fulfilled through "community outreach events like open houses, free lectures and events, and STEM education opportunities" and that scientists can also "attend community events, visit local schools, and much more."

DOE contracts with the national labs also charge them with public engagement. For example, Argonne's contract states that laboratory management is responsible for "the development, planning, and coordination of proactive approaches for the timely dissemination of unclassified information regarding DOE activities onsite and offsite, including, but not limited to, operations and programs." Examples of ways to do this include "public workshops, meetings or hearings, open houses, newsletters, press releases, conferences, audio/visual presentations, speeches, forums, tours, and other appropriate stakeholder interactions."

For all groups at a national lab, enhanced visibility to the public is a desirable situation, and the laboratory should both facilitate this communication and reward those individuals who choose to spend some of their time on this important goal.

**Why national labs are influential in enabling scientists to participate in public engagement:**
Just like colleges and universities, national labs have an important role to play in setting good conditions for public engagement activities to thrive. The rubrics for hiring and promotion within national labs help set the priorities of the scientists and users, and therefore the directions of the research groups. If public engagement were expected, supported and rewarded, the culture within national labs would respond, and in turn the culture of particle physics would change.

Fermi National Accelerator Laboratory recently launched a new process to allow laboratory employees to charge for time spent on either equity, diversity and inclusion-related work (EDI) or education and public engagement (EPE). According to an article about the change: "Under the new procedure, employees working 40+ hours a week may charge up to six hours per month on EDI/EPE combined without seeking prior manager approval." They may request manager approval for anything above six hours. The article states that "lab leadership encourages participation in EDI and EPE."

Senior management sets the tone for the entire laboratory. If senior management values and enables public engagement, middle management will do the same, and all individuals at the laboratory will be empowered to participate.

**Recommendations:**
- Provide training in public engagement.
- Provide opportunities for employees and users to participate in public engagement activities.
- Identify individuals with talent and interest in public engagement and consider this effort to be part of their assigned duties.
- Recognize the value of public engagement activities and offer compensatory time off for such efforts.
- Prior to making an offer in hiring decisions, require a statement from hiring managers on selected candidates' outreach achievements and potential.
- Require that all promotion packages have a statement on employee's outreach efforts and achievements.
- Include outreach in annual merit reviews.
- Include an outreach category for annual performance awards, in line with other achievements, such as managerial, technical, and publication achievements.
- Lay out expectations for public engagement in staff handbooks.

# 2.4 Government departments and policy makers
## 2.4.1 Office of Science and Technology Policy

**Background:**
The Office of Science and Technology Policy advises the executive branch and leads the process of developing US Science and Technology strategies. Established in 1976, OSTP includes in its scope the "scientific, engineering, and technological aspects of the economy, national security, homeland security, health, foreign relations, and the environment."

**How public engagement helps OSTP achieve their goals:**
The mission of OSTP is to "maximize the benefits of science and technology to advance health, prosperity, security, environmental quality, and justice for all Americans." Public engagement can raise the profile of scientific results and help excite the public around ongoing scientific research. Stronger support of science among the public can strengthen the funding profile for science, and help maximize the impact of this research, consistent with the mission of OSTP.

**How OSTP is influential in enabling scientists to participate in public engagement:**
OSTP leads the process of developing US science and technology policies. They work across many sectors, such as philanthropy, local governments, and academic communities. By setting priorities they can have a direct influence on the US budget.

**Recommendations:**
If OSTP recognizes the link between (1) public engagement of scientists with the public and (2) potential support for and benefits of scientific research to the public, they can serve as real drivers of progress. This recognition could manifest itself in:
- strategies that are crafted to include public engagement as a priority in action plans, and
- the inclusion of public engagement in prioritization memos written by OSTP.

## 2.4.2 Congress

**How public engagement helps Congress achieve their goals:**
Government-supported research is a crucial component within the ecosystem of science funding in the US. Strong public engagement by scientists can make clear the link between scientific research and development and the security and prosperity of the country.

**How Congress is influential in enabling scientists to participate in public engagement:**
Congress, which controls the government purse strings, plays an important role in the support of an ecosystem of public engagement.

**Recommendations:**
- Expand funding for the National Science Foundation and the Office of Science in the Department of Energy, which will create space in those budgets for public engagement activities.
- Provide sufficient funds for workforce development, which can support training programs that include public engagement from scientists.
- Maintain a focus on diversifying the workforce, which can help support public engagement programs that are targeted to reach groups that are under-represented in our field.

# 2.5 Institutions that fund research
## 2.5.1 Department of Energy

**Background:**
The Department of Energy is central to the funding of large, international projects in high-energy physics. DOE's mission does not specifically mention public engagement, but public engagement activities can help advance DOE's goal of developing a diverse workforce in STEM.

A Committee of Visitors periodically reviews the program offices in the Department of Energy. Both 2016 and 2020 Committee of Visitors reports advised DOE HEP to support efforts to improve the diversity of the field.

In 2016, the Committee of Visitors wrote in their recommendations that "[u]nder-representation of women and minorities in physics as a whole continues to be a challenge. Greater attention should be paid to promoting an inclusive environment in order to provide encouragement to research groups to improve the diversity of the HEP workforce. HEP review processes for university groups and laboratories should consider activities that promote diversity and inclusion in the workforce and the workplace."

In 2020, the Committee of Visitors complimented HEP staff in its report "for their considerable effort to address diversity, equity, and inclusion (DEI)," writing that they "express significant awareness of the issues and their importance." The committee recommended HEP "develop and implement strategies and policies to foster diversity, equity, and inclusion in supported university groups as well as at the laboratories." The 2020 Committee of Visitors report also commended HEP for adding "mentorship and other diversity considerations" to the "Program Policy Factors" that add to the strength of a grant proposal.

The Department of Energy Office of Science has two offices committed to promoting the development of a diverse STEM workforce at the US national laboratories and in the United States in general.

The Office of Science's Office of Workforce Development for Teachers and Students "sponsor[s] workforce training programs designed to motivate students and educators to pursue careers that will contribute to the Office of Science's mission in discovery science and science for the national need." According to the Office of Science website: "The mission of [WDTS] is to help ensure that DOE and the Nation have a sustained pipeline of highly skilled and diverse science, technology, engineering, and mathematics (STEM) workers."

The Office of Science's Office of Scientific Workforce Diversity, Equity, and Inclusion "collaborates across [the Office of Science] to advance organizational best practices…to promote DEI at the SC DOE National Laboratories."

DOE sets expectations for laboratory involvement in public engagement in their contracts, both for the purposes of "[building] a positive, supportive relationship founded on openness and trust with the community and region in which the Laboratory is located" and for "the timely dissemination of unclassified information regarding DOE activities onsite and offsite, including, but not limited to, operations and programs." (For example, see Argonne's contract.) According to the DOE article "Community, Science, and Your Neighbor," examples of ways to do this include "community outreach events like open houses, free lectures and events, and STEM education opportunities," in addition to community events and visits to local schools.

Through a partnership with the Kavli Foundation, DOE has set up SciPEP, a partnership with a goal of "ensur[ing] scientists are supported to be effective communicators and, if appropriate, active in engaging the public."

**How public engagement helps DOE achieve their goals:**
Public engagement specifically designed to reach marginalized communities lines up well with DOE's goal of developing a diverse STEM workforce. The Office of Science specifically mentions public engagement in its "Summary of DOE Office of Science Recognized Promising Practices for DOE National Laboratory Diversity, Equity, and Inclusion Efforts," which recommends "supporting general laboratory educational outreach and community involvement and outreach activities" that are linked to both "the laboratory mission, goals and objectives" and "DEI strategic goals and objectives."

**Why DOE is influential in enabling scientists to participate in public engagement:**
Funding agencies play an extremely important role in setting researchers' priorities. If public engagement helps a scientist receive research funding, not only will the scientist prioritize it; leadership at their institution will prioritize it as well.

**Recommendations:**
- Add public engagement to merit review criteria.
- Financially support efforts to provide scientists training in public engagement.
- Financially support efforts to create public engagement programs specifically designed to reach marginalized groups.
- Continue to set expectations in laboratory contracts that the national laboratories participate in public engagement.
- Include funding in support of the Community Engagement Frontier as a part of the Snowmass process. This should include support for workshops that provide training and strategy development for public engagement in particle physics.

# 2.5.2 National Science Foundation

**Background:**
Scientists submitting research proposals to the National Science Foundation must include a statement explaining the "broader impacts" of their research, which the NSF Proposal and Award Policies & Procedures Guide defines as "the advancement of scientific knowledge and activities that contribute to the achievement of societally relevant outcomes." These "broader impacts" can include:

- Full participation of women, persons with disabilities and underrepresented minorities in STEM
- Improved STEM education and educator development at any level
- Increased public scientific literacy and public engagement with science and technology
- Improved well-being of individuals in society

- Development of a diverse, globally competitive STEM workforce
- Increased partnerships between academia, industry and others
- Improved national security
- Increased economic competitiveness of the US
- Enhanced infrastructure for research and education

Specifically to support public engagement efforts, NSF also runs the [Advancing Informal STEM Learning](#) (AISL) program, which "seeks to advance new approaches to and evidence-based understanding of the design and development of STEM learning opportunities for the public in informal environments; provide multiple pathways for broadening access to and engagement in STEM learning experiences; advance innovative research on and assessment of STEM learning in informal environments; and engage the public of all ages in learning STEM in informal environments." The program is supported by the [Center for the Advancement of Informal Science Education](#).

**How public engagement helps NSF achieve their goals:**
NSF's Social, Behavioral and Economic Sciences Directorate published a "[Dear Colleague](#)" Letter naming two ways that research can create "broader impacts": (1) through scientific opportunities, and—most relevant to this paper—(2) through communicative products, "the vehicles by which researchers share their findings…with other researchers, policymakers, and the public."

**Why NSF is influential in enabling scientists to participate in public engagement:**
Explicitly naming communication with the public as a way to fulfill a requirement of the grant application process has encouraged scientists to participate in public engagement.

**Recommendations:**
- Continue to require a broader impacts statement in merit reviews.
- Continue to award grants to scientists who include public engagement plans in their grant proposals.

# 2.5.3 Private foundations

**Background:**
The private foundation most involved with public engagement by physicists is the [Kavli Foundation](#). Its mission is to "advance science for the benefit of humanity," with a special focus on "basic research in the fields of astrophysics, nanoscience, neuroscience and theoretical physics."

The Kavli Foundation's Science in Society program "go[es] beyond scientific research and focus[es] on ways to strengthen the relationship between science and society;" this includes the focus area "[Public Engagement with Science](#)." Through a partnership with the Department of Energy, the Kavli Foundation has set up [SciPEP](#), a partnership with a goal of "ensur[ing] scientists are supported to be effective communicators and, if appropriate, active in engaging the public." They also support scientists' participation in public engagement through their Civic Science Fellowship, the Leaders in Science and Technology Engagement Networks, the Network for Science Communication Trainers, and the Alda-Kavli Learning Center.

Other foundations support scientists in specific public engagement activities. The Sloan Foundation's strategy related to Public Understanding of Science, Technology & Economics "focuses on books, theater, film, television, radio, and new media." The Simons Foundation offers support to organizations that do public engagement, such as the World Science Festival Foundation.

**How public engagement helps private foundations achieve their goals:**
Private foundations are interested in having a large impact through their work. Public engagement is a critical piece of that impact. By supporting public engagement, not only can foundations support research that will further our understanding of the universe, they can also share that understanding with the widest possible audience.

Private foundations are also interested in being recognized for their work, and public engagement can make a larger number of people aware of the foundations' investments.

**Why private foundations are influential in enabling scientists to participate in public engagement:**
Private foundations have resources that allow them to influence the agendas of practicing scientists.

**Recommendations:**
- Ensure that a percentage of funding goes toward public engagement.
- Explicitly make calls for public engagement proposals.
- Fund public engagement training for scientists.

The particle physics community can take an active role in forming partnerships with private foundations to encourage the development of new programs or collaborations that are aligned with public engagement.

# 2.6 Professional bodies and societies
## 2.6.1 American Association for the Advancement of Science and American Physical Society

**Background:**
The American Association for the Advancement of Science values public engagement. Its AAAS Center for Public Engagement with Science seeks to "[provide] scientists and scientific institutions with opportunities and resources to have meaningful conversations with the public." It "focuses on connecting research about science communication and public engagement—what works and what doesn't, the benefits of engagement, and why scientists engage—with scientists and practitioners who can put research into practice." It accomplishes this with a variety of programs.

The center also gives awards for public engagement: the AAAS Mani L. Bhaumik Award for Public Engagement with Science and the AAAS Early Career Award for Public Engagement with Science, both of which come with a monetary award of $5,000.

The American Physical Society also supports public engagement in a variety of ways. For one, APS has a 9-member Committee on Informing the Public, which "provides oversight of the Society's public outreach and media relations activities while also seeking mechanisms to encourage or facilitate public outreach by APS members." Members of the committee "suggest future activities, approaches, and outreach opportunities, as well as possible external funding sources" and also make budget recommendations to APS.

APS members have also organized a Forum on Outreach & Engaging the Public, whose "goal is to increase the public's awareness of physics by providing a forum with APS for the large number of physicists currently involved in a diverse array of outreach and public engagement activities." They recognize public engagement efforts with the Dwight Nicholson Medal for Outreach, which includes a $3,000 stipend, and by nominating APS members involved in public engagement for the APS Fellowship.

APS hosts a series of webinars related to "Engaging the Public through Science." They also award "mini grants" of up to $10,000 to support public engagement activities. (For examples of projects funded by APS Outreach Mini Grants, see a 2019 article in APS News.)

An APS committee recently proposed an APS statement "that encourages academic, research, and other institutions to add the participation in informal science education activities to the criteria they use for hiring and career advancement decisions." Among the benefits the paper lists of these activities are "enhancing the critical thinking, understanding of public policy, career opportunities, and science literacy of the public; improved communication skills, methodology, and mentoring by researchers; increased science enrollment, visibility, reputation, and research funding for institutions; and expanded and more diverse recruiting and more public support for the field of physics."

**How public engagement helps AAAS and APS achieve their goals:**
The American Association for the Advancement of Science website lists nine "Focus Areas," including "Public Engagement," which it describes as "intentional, meaningful interactions that provide opportunities for mutual learning between scientists and the public." The site states:

"Goals for public engagement with science include civic engagement skills and empowerment, increased awareness of the cultural relevance of science, and recognition of the importance of multiple perspectives and domains of knowledge to scientific endeavors. In order to 'advance science and serve society,' AAAS engages the public and empowers scientists to engage the public on issues related to research, education, policy, and more."

The American Physical Society specifically mentions public outreach in its 2019 Strategic Plan. Its mission includes "shar[ing] the excitement of physics and communicat[ing] the essential role physics plays in the modern world." Its vision includes "expand[ing] public appreciation of physics and its many contributions."

The APS Strategic Plan states: "In order to be the leading voice for physics in the US, we will support member engagement in effective…public outreach," specifying that APS will do this by "helping members become effective and informed advocates for science." Further, the strategy states: "In order for APS to continue to play a leadership role in innovative and impactful science education, outreach, and diversity programs, we will ensure that these programs…are well-positioned to receive robust support and recognition." The plan specifies that APS will do this by "exploring better ways to communicate" and "offering opportunities for members to participate in public engagement."

**Why AAAS and APS are influential in enabling scientists to participate in public engagement:** The majority of particle physicists are associated with professional bodies and societies. Professional bodies and societies register members, organize meetings and conferences, implement ethics codes, and make statements on behalf of their members. They have close connections with other structures such as physics departments at universities and colleges.

It is very important that the professional societies are leading from the front so that individual particle physicists find the courage to initiate new public engagement programs or join existing ones. AAAS and APS have done an excellent job of providing a wide range of training related to public engagement. We applaud both professional societies for their recognition of public engagement efforts and for organizing public engagement around conferences.

**Recommendations**:
- Continue providing public engagement training.
- Continue organizing public engagement activities in conjunction with scientific conferences.
- Recommend conference hosts include in their proposals information about how they will support public engagement—for large conferences, require it.
- Recommend conference hosts provide public engagement training in their programming —for large conferences, require it.
- Recommend conference hosts involve local public affairs professionals in training and public engagement planning.
- Prioritize activities that will allow the largest number of physicists to participate.
- Continue honoring scientists who participate in public engagement with awards.

- Continue supporting research into effective public engagement.
- Continue to codify these programs in APS and AAAS strategies and planning documents.
- Help members push for recognition of and support for public engagement efforts at all levels, from funding agencies through university departments.

# 2.6.2 APS Division of Particles & Fields

**Background:**
The APS Division of Particles & Fields organizes the Snowmass process to inform the Particle Physics Project Prioritization Plan over the forthcoming decade.

DPF holds a conference every other year. Going back to at least 2009, DPF conferences have always included parallel sessions related to "Outreach and Education" (in one case "Diversity, Education, and Outreach"). Each DPF conference from 2009 through 2020 has included a public lecture. (In 2021, the DPF conference was virtual.) In 2011 and 2013, DPF conferences included lunchtime discussions related to communication—one on "Physics and Modern Media," organized by a science journalist, and another on "Outreach Activities." In 2009, 2011, 2017, 2019 and 2021, the DPF conferences included plenary sessions related to Outreach and Education.

The DPF website includes a list of links collected on a [resource page](#) for members interested in Education and Outreach.

**How public engagement helps DPF achieve their goals:**
Professional bodies and societies seek to advocate on behalf of scientists and the sciences. An important part of that effort is empowering their members to engage the public.

**Why DPF is influential in enabling scientists to participate in public engagement:**
Professional organizations play an important role in shaping the culture of particle physics. If public engagement is important in professional organizations, physicists will be encouraged to prioritize it as well.

**Recommendations**:
- Take full ownership of the public engagement recommendations and institutionalize this responsibility within the DPF leadership structure.
- Continue to include the Community Engagement Frontier in the Snowmass process.
- Advocate for funding for Community Engagement Frontier activities within the Snowmass process.
- Provide public engagement training.
- Continue offering public lectures in conjunction with scientific conferences.
- Organize public engagement activities for multiple scientists to participate in conjunction with scientific conferences, including prioritization processes like Snowmass.

- Recommend conference hosts include in their proposals information about how they will support public engagement.
- Recommend conference hosts provide public engagement training in their programming.
- Recommend conference hosts involve local public affairs professionals in training and public engagement planning.
- Honor scientists who participate in public engagement with awards.
- Codify the inclusion of public engagement plenary sessions and parallel sessions in DPF conferences.
- Help members push for recognition of and support for public engagement efforts at all levels, from funding agencies through university departments.
- Encourage DPF members to participate in APS public engagement training programs and activities.

# 3. Conclusion

In the process of surveying colleagues and having conversations with various constituencies throughout the Snowmass process, we have witnessed the energy and creativity that particle physicists demonstrate In their engagement with the public. And we have learned that much of this ongoing work is happening in spite of structural barriers that physicists face. This widespread lack of support for work that is vital to the success of our field prompted us to think across the many institutions and groups within which we operate to identify what structural changes are needed for a true cultural shift toward encouragement of high-quality public engagement.

In this document we have recommended structural changes that will encourage and support the work of public engagement with science by particle physicists. Some of these recommendations could be adopted immediately, whereas others may take some time or face roadblocks that cannot easily be cleared. No one group or individual has the power to implement all of these changes, but we each have a role that we can play to improve the situation. We therefore advocate for forward progress in all of these areas simultaneously. We encourage each of our colleagues, and particularly those with leadership positions, to think about what influence they have to move items of this agenda forward. With many individuals working toward change on many axes, we can begin the important work of changing the culture within, and surrounding, our field.

When the P5 process is completed and priorities for our field have been established, we should consider a strategy for public engagement that will amplify our findings and support the adopted agenda. Each of us has a role to play, either from within a research group, university, laboratory, professional society, funding agency or from within the government.

We recommend that future Snowmass processes include a group dedicated to public engagement, and that the work of this group include an assessment of where we stand with respect to the structural changes advocated for in this document. As we recommend above, DPF has an important role to play in tracking progress and taking responsibility for continuing to push forward the work outlined in this paper.